\documentclass{article}
\usepackage{spconf,amsmath,graphicx}
\usepackage{hyperref}
\usepackage[utf8]{inputenc}
\usepackage[T1]{fontenc}
\usepackage{color,soul}
\usepackage{times}
\usepackage{subfigure}
\usepackage{xcolor}
\usepackage{epsfig}
\usepackage{graphicx}
\usepackage{amssymb}
\usepackage{multicol,multirow}
\usepackage{booktabs}
\usepackage{bigstrut}
\usepackage{floatrow}
\usepackage{rotating}
\usepackage{adjustbox}
\usepackage{amsmath}
\usepackage{longtable}
\usepackage{array}
\def\x{{\mathbf x}}

\title{LMBF-Net: A Lightweight Multipath Bidirectional Focal Attention Network for Multifeatures Segmentation}
\name{Tariq M Khan$^1$, Shahzaib Iqbal\,$^2$, Syed S. Naqvi\,$^3$, Imran Razzak$^1$, Erik Meijering$^1$}
\address{$^1$School of Computer Science and Engineering, University of New South Wales, Sydney, NSW, Australia\\$^2$Department of Electrical Engineering, Abasyn University, Islamabad,\\ $^3$Department of ECE, COMSATS University Islamabad (CUI), Islamabad, Pakistan}
\begin{document}
%
\maketitle
\begin{abstract}
Retinal diseases can cause irreversible vision loss in both eyes if not diagnosed and treated early. Since retinal diseases are so complicated, retinal imaging is likely to show two or more abnormalities. Current deep learning techniques for segmenting retinal images with many labels and attributes have poor detection accuracy and generalisability. This paper presents a multipath convolutional neural network for multifeature segmentation. The proposed network is lightweight and spatially sensitive to information. A patch-based implementation is used to extract local image features, and focal modulation attention blocks are incorporated between the encoder and the decoder for improved segmentation. Filter optimisation is used to prevent filter overlaps and speed up model convergence. A combination of convolution operations and group convolution operations is used to reduce computational costs. This is the first robust and generalisable network capable of segmenting multiple features of fundus images (including retinal vessels, microaneurysms, optic discs, haemorrhages, hard exudates, and soft exudates). The results of our experimental evaluation on more than ten publicly available datasets with multiple features show that the proposed network outperforms recent networks despite having a small number of learnable parameters.
\end{abstract}
\section{Introduction}
\label{sec:intro}

Ophthalmologists rely on retinal image analysis to diagnose a variety of vision-threatening disorders, including diabetic retinopathy (DR), glaucoma, and age-related macular degeneration (AMD). DR is one of the leading causes of blindness \cite{iqbal2022recent}, and characteristics such as slope, neovascularisation, and curvature of retinal blood vessels are important in the diagnosis of such retinal diseases \cite{franklin2014computerized,iqbal2023robust}. Pathologies commonly observed in diseased retinas include microaneurysms, hard and soft exudates, and hemorrhages \cite{manan2023semantic}. Medical image segmentation is crucial to understanding image details and pinpointing areas of injury or anomaly\cite{abbasi2023lmbis, khalid2023advancing}. This process forms the bedrock for a multitude of medical image analysis techniques, including, but not limited to, 3D reconstruction and registration \cite{khan2023retinal}. Its importance reverberates through clinical settings, profoundly influencing the accuracy of the diagnosis and subsequent treatment outcomes. The advent of deep learning has contributed significantly to medical image analysis, particularly in segmentation. Traditional models for semantic segmentation, such as FCN \cite{long2015fully} and U-shaped networks (U-Net), typically focus on details at the pixel level \cite{chen2021transunet, khan2024esdmr}.

The key components of the mentioned segmentation architectures are encoders and decoders. Medical image segmentation architectures typically use the most prominent feature extractors, such as ResNet \cite{gu2019net, iqbal2023ldmres} as encoders. Currently, multiscale pooling, dilated convolution, and other techniques are used to create semantic feature extraction modules \cite{iqbal2022g, naqvi2023glan, khan2022mkis}. The attention mechanism is frequently used to extract semantic information to effectively focus on semantic aspects \cite{li2021pyconvu,iqbal2023mlr}. Self-attention (SA) is considered crucial to its success, as it allows global input-dependent interactions, distinguishing it from convolution operations that confine interactions to local regions with shared kernels \cite{liu2021swin}. However, despite its advantages, the efficiency of SA has raised concerns due to its complexity with respect to visual token numbers, especially in scenarios with high-resolution input \cite{yang2022focal}. Focal modulation \cite{yang2022focal} is a recent attention mechanism that has several benefits over SA mechanisms. It is more efficient than SA as it does not require the computation of pairwise attention weights, making it the best choice for lightweight architectures. In addition, U-Net is often used as the backbone for improved medical image segmentation models \cite{khan2020semantically, mehmood2024retinalitenet, khan2023feature}. Skip connections are used by U-Net to efficiently increase low-level features, although they result in information redundancy \cite{khan2022t}. 

A bidirectional recurrent O-shaped U-Net named Bio-Net \cite{xiang2020bio} recursively maps the decoder features back to the encoder through skip connections. Bio-Net is computationally expensive for binary segmentation problems, as it uses standard convolution operations and takes a lot of time for convergence due to filter overlapping. The basic design of Bio-Net causes the images to become blurry at the bottleneck layers of the network. Thus, there is a need to refine the encoder features at the bottleneck layers. Furthermore, the segmentation speed is important for clinical applications. In various clinical scenarios such as live real-time disease monitoring, image-guided surgery, and adaptive radiation therapy, the complexity of U-Net-based models poses limitations on segmentation performance. From a practical perspective, there is a need for less complex networks that are computationally more efficient, without sacrificing segmentation accuracy.

To solve these problems, a novel approach for medical image segmentation is proposed, introducing the Lightweight Multipath Bidirectional Focal Attention Network (LMBF-Net) with a remarkably low number of learnable parameters (0.191M). By integrating a blend of convolution and group convolution operations, the proposed LMBF-Net demonstrates substantial advances in computational efficiency. Through optimization of filter numbers, the network mitigates filter overlap, ensuring expedited convergence compared to its predecessor, Bio-Net, thus significantly reducing training duration. Crucially, bidirectional skip connections facilitate seamless information exchange between the encoder and decoder, fostering enhanced feature extraction. Moreover, the integration of a Focal Modulation Attention Block (FMAB) between the encoder and decoder elevates the quality of the encoder features. Addressing the challenge of class imbalance in medical image segmentation, we introduce a patch selection strategy tailored to enhance the performance of LMBF-Net. Extensive evaluation of various features of retinal images, including retinal vessels, hard exudates, soft exudates, haemorrhages, microaneurysms, and the optic disc, underscores the robustness and generalisability of the proposed LMBF-Net.

\section{Proposed LMBF-Net}
\label{sec:MB Net}

\subsection{Network Structure}
LMBF-Net uses the U-Net-type encoder-decoder structure as the backbone (Fig.~1). Traditional encoder-decoder structures usually have a single-scale input. The proposed LMBF-Net starts from a single path to extract the image features first and then represents the image features at different scales using a Multipath Residual Block (MRB). The single path starts with the 256$\times$256 input layer with 3 channels for RGB images followed by the 3$\times$3 convolutional layer with 8 channels. The rectified linear unit (ReLU) activation layer and batch normalisation (BN) layers are used after the first convolutional layer. 
The subsequent first MRB has 16 channels with 4 different paths. After that, the first downsampling is performed by using a max-pooling layer. Once the first downsampling is performed, the downsampled features are fed into the second 32-channel MRB followed by the second downsampling. The LMBF-Net uses only three max-pooling layers to minimise spatial information loss. Once all three max-pooling layers are employed, the proposed LMBF-Net refines the encoded features and enhances the channel mapping by applying FMAB. The information extracted improves the sensitivity and specificity of the proposed LMBF-Net. Once the encoder features are refined by focal modulation, a transpose convolutional layer performs the first upsampling and is followed by the 32 channels MRB, the second upsampling layer, and the 16-channel MRB. In the end, a 3$\times$3 convolutional layer with 8 channels is used, and the final output segmentation maps are generated by a softmax layer followed by a Dice pixel classification layer. The proposed network integrates bidirectional skip connections across its encoding and decoding layers, enabling the fusion of low- and high-level feature information through forward skip connections. In addition, reverse skip connections are used to map the decoded characteristics back to the encoder.


\begin{figure}[!t]
  \centering
  \resizebox{1\columnwidth}{!}{%
\includegraphics[width=1\columnwidth]{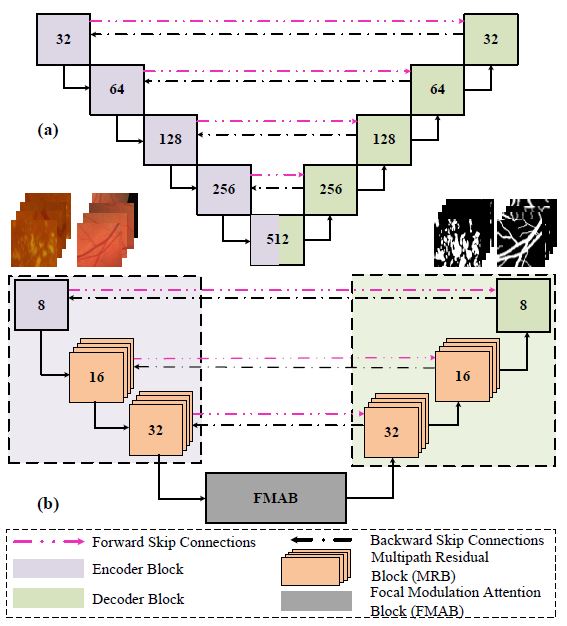} 
  }
  \label{fig:model}
\caption{\small The proposed LMBF-Net architecture. (a) Bio-Net architecture \cite{xiang2020bio}. (b) LMBF-Net architecture.}

\end{figure}%

\begin{figure*}[!t]
  \centering
  \resizebox{1\columnwidth}{!}{%
\centering
\subfigure[]{\includegraphics[width=1\columnwidth]{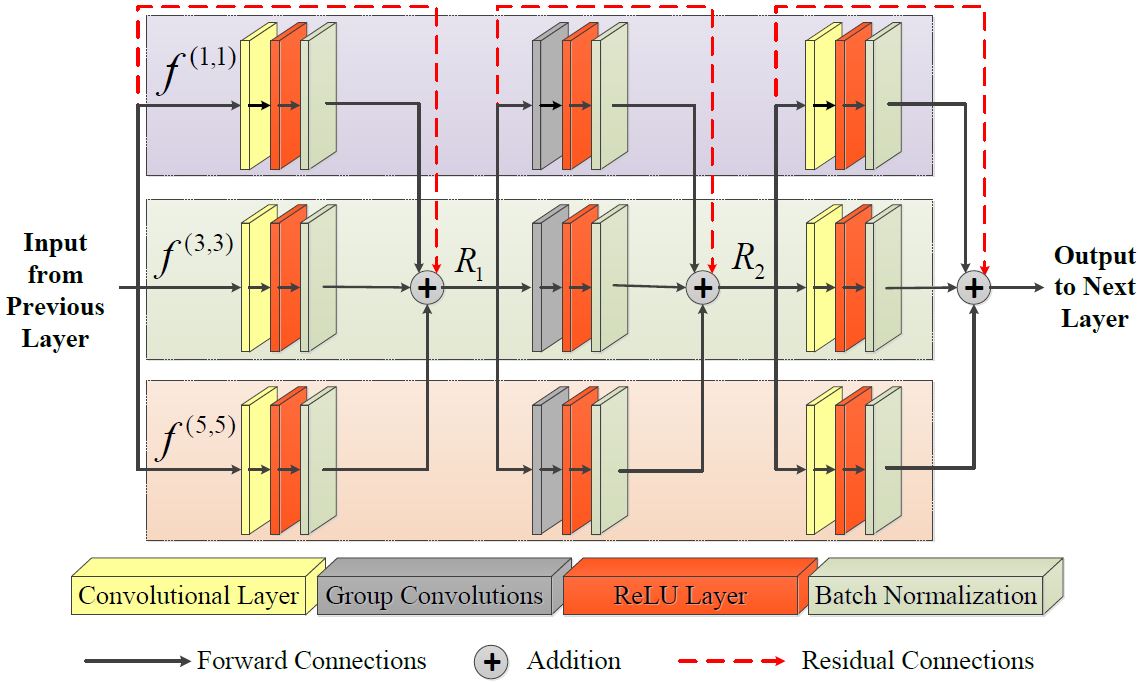}}
  \centering
\subfigure[]{\includegraphics[width=1\columnwidth]{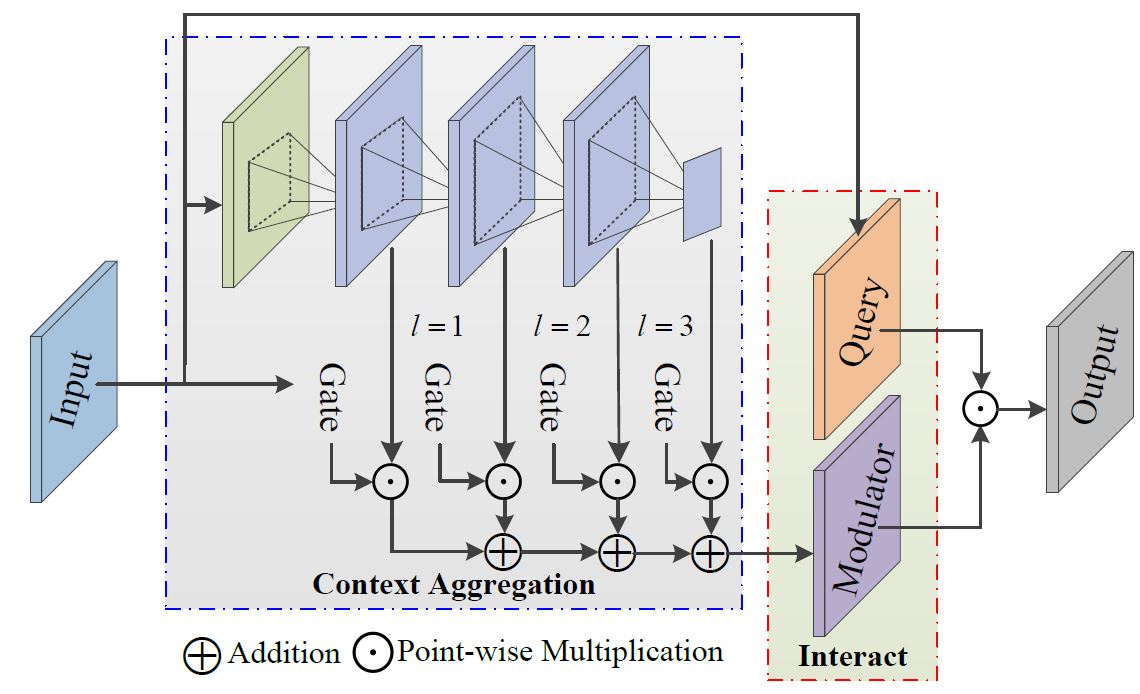} }
  }
  \label{FMB}
\caption{\small The proposed LMBF-Net architecture. (a) Multipath Residual Block (MRB). (b) Focal Modulation Attention Block (FMAB).}

\end{figure*}%

\subsection{Multipath Residual Block (MRB)}
The MRB is used to extract multiscale features at different kernel scales and then adds them to combine these features (Fig. 2a). It starts from the three parallel convolution paths of 1$\times$1, 3$\times$3, and 5$\times$5 followed by the ReLU activation layer and batch normalisation. These parallel paths are then added along with the input features. The result of the addition is passed through the group convolutions of 1$\times$1, 3$\times$3, and 5$\times$5 followed by the activation layer ReLU and BN, and then the addition of these connections is carried out along with the output of the first addition. Finally, the output of the second addition is fed to another multipath block of three parallel convolution paths of 1$\times$1, 3$\times$3 and 5$\times$5 followed by the ReLU activation layer and BN. The final output of the MRB is computed by adding these paths along with the output of the second addition.

Let $\Im_\text{in}$ be the input feature map fed to the MRB. $R_1$ is the first addition of the multipath features and residual connection, computed as:
\begin{equation}
    R_1=\Im_\text{in}+\sum_{i=1}^{3}\text{BN}(\sigma(f_\text{c}^{n\times n}(\Im _\text{in}))),
\label{Eq:MB1}
\end{equation}
where $n$ represents the size of the convolutional kernel and is computed as $n=2i-1$, $f_\text{c}^{n\times n}$ is the convolution operation of size ${n \times n}$, ${\sigma}$ denotes the ReLU activation function and BN the batch normalisation. $R_2$ is the second addition of the multipath features and $R_1$, and is computed as:
\begin{equation}
    R_2=R_1+\sum_{i=1}^{3}\text{BN}(\sigma(f^{n\times n}_\text{g}(R_1))),
\label{Eq:MB2}
\end{equation}
where ${f^{n\times n}_\text{g}}$ denotes the group convolution operation of size $n \times n$. Finally, the output ${\Im_\text{out}}$ of the MRB is computed by adding the multipath features with $R_2$:
\begin{equation}
    \Im _\text{out}=R_2+\sum_{i=1}^{3}\text{BN}(\sigma(f^{n\times n}(R_2))).
\label{Eq:MB3}
\end{equation}

Three residual blocks are cascaded in each multipath block. The first residual block is used to standardise and extract features from multiple scales and depths. The second residual block is responsible for extracting fine-grained feature information from each scale. The idea is to learn scale-aware features to capture and refine objects at each scale. This is achieved through grouped convolution, where our aim is to learn a mapping between scales and filter groups. The grouped convolution also eliminates redundant features, thus ensuring separation between scales. The final residual block is responsible for aggregating the feature information from multiple scales, sizes, and depths. The cascading of residual blocks and the skip connections enable global information propagation, as well as mid- and high-level frequency information propagation to later layers.

\subsection{Focal Modulation Attention Block (FMAB)}
The FMAB \cite{yang2022focal} is used between the encoder and the decoder to further enhance the extracted feature information. It is composed of three distinct components (Fig.2b). First, the FMAB takes input from the last layer of the encoder block and then uses a stack of depth-wise convolutional layers to encode visual contexts, spanning from short to long ranges. Each layer of the stack captures various levels of granularity, encompassing local and global information. Next, this extracted information is selectively aggregated into context features for each query token, guided by its content. A learnable attention weight serves as a gate to determine the importance of each context feature in the final representation of the query token. Finally, these aggregated context features are fused into the query token using an element-wise affine transformation. This transformation is defined by a learnable weight matrix that is updated during the training process. 

\subsection{Implementational Details}
All network training was performed on Google Colab with an NVIDIA K80 GPU with 16 GB RAM. Keras was used as the deep learning framework. The initial learning rate was set as 1e-3. The number of iterations for each epoch was set to 250. Finally, Dice-pixel loss was used at the training stage. LMBF-Net was trained using Adam optimiser for 150 epochs with batch sizes of 16 for vessel segmentation and 4 for DR lesion segmentation, respectively.

\section{Experiments and Results}
\label{sec:Results}

LMBF-Net was compared to various modern segmentation networks. Here, we present an overview of the used datasets, patch generation and selection strategy, and evaluation criteria for segmentation, followed by the results of an ablation study and the comparative experiments.

\begin{table}[!b]
  \centering
  \adjustbox{max width=\textwidth}{%
    \begin{tabular}{llccccl}
    \toprule
    \multirow{2}[4]{*}{\textbf{Dataset}} & \multirow{2}[4]{*}{\textbf{Feature}} & \multicolumn{3}{c}{\textbf{Number of Images}} & \multirow{2}[4]{*}{\textbf{Format}} & \multirow{2}[4]{*}{\textbf{Online Source}} \\
    \cmidrule{3-5} & & \textbf{Training} & \textbf{Testing} & \textbf{Total} & & \\
    \midrule
    DRIVE \cite{DRIVEdata} & \multirow{4}[1]{*}{Blood Vessels} & 20    & 20    & 40   & TIFF  & \href{https://www.kaggle.com/datasets/andrewmvd/drive-digital-retinal-images-for-vessel-extraction/}{Kaggle} \\
    STARE \cite{STAREDataset}&       & 10    & 10     & 20    & PPM  & 
    \href{https://cecas.clemson.edu/~ahoover/stare/}{Clemson University} \\
    CHASE \cite{CHASEDataset}  &       & 20    & 8     & 28    & JPEG  & 
    \href{https://blogs.kingston.ac.uk/retinal/chasedb1/}{Kingston University London} \\
    HRF \cite{HRFDataset}  &       & 23    & 22     & 45    &  JPEG     & 
    \href{https://www5.cs.fau.de/research/data/fundus-images/}{Friedrich Alexander University} \\
    \midrule
    \multirow{5}[1]{*}{IDRiD \cite{IDRiDDataset} } & Microaneurysms & \multirow{5}[1]{*}{54} & \multirow{5}[1]{*}{27} & \multirow{5}[1]{*}{81} & \multirow{5}[1]{*}{JPEG} & \multirow{5}[1]{*}{\href{https://ieee-dataport.org/open-access/indian-diabetic-retinopathy-image-dataset-idrid}{IEEE DataPort}} \\
          & Haemorrhages &       &             &       &       &  \\
          & Hard Exudates &       &             &       &       &  \\
          & Soft Exudates &       &             &       &       &  \\
          & Optic Disc &       &            &       &       &  \\
    \bottomrule
    \end{tabular}}
    \vspace{-1em}
  \caption{Retinal image datasets used for the experiments.}
  \label{tab:Datasets}%
\end{table}%

\begin{table}[!b]
  \centering
    \adjustbox{max width=\textwidth}{
    \begin{tabular}{lccclcccc}
    \toprule
    \multirow{3}[6]{*}{\textbf{Dataset}} & \multicolumn{1}{c}{\textbf{Orignal}} & \multicolumn{1}{c}{\textbf{Resized }} & \multirow{3}[6]{*}{\textbf{Patch Size}} & \multirow{3}[6]{*}{\textbf{Feature}} & \multicolumn{4}{c}{\textbf{Training Details}} \\
\cmidrule{6-9}          & \multicolumn{1}{c}{\textbf{Image }} & \multicolumn{1}{c}{\textbf{Image }} &       &       & \multirow{2}[4]{*}{\textbf{Images}} & \multicolumn{3}{c}{\textbf{Patches}} \\
\cmidrule{7-9}          & \multicolumn{1}{c}{\textbf{Size}} & \multicolumn{1}{c}{\textbf{Size}} &       &       &       & \textbf{Total} & \textbf{Selected} & \textbf{Discarded} \\
    \midrule
    DRIVE \cite{DRIVEdata} & 565 $\times$ 584 & 640 $\times$ 640  & \multirow{4}[2]{*}{128 $\times$ 128} & \multirow{4}[2]{*}{Blood Vessels} & 20    & 500   & 500   & - \\
    STARE \cite{STAREDataset} & 605 $\times$ 700 & 640 $\times$ 640  &       &       & 10    & 250   & 250   & - \\
    CHASE \cite{CHASEDataset} & 960 $\times$ 990 & 1,025 $\times$ 1,024 &       &       & 20    & 1,280  & 1,280  & - \\
    HRF \cite{HRFDataset}   & 3,504 $\times$ 2,336 & 3,456 $\times$ 2,304 &       &       & 23    & 11,178  & 11,178  & - \\
    \midrule
    \multirow{5}[2]{*}{IDRiD \cite{IDRiDDataset}} & \multirow{5}[2]{*}{2,848 $\times$ 4,288} & \multirow{5}[2]{*}{$2,816 \times 4,096$} & \multirow{4}[1]{*}{$256 \times 256$} & Hard Exudates & 54    & 9,504  & 1,680  & 7,824 \\
          &       &       &       & Soft Exudates & 26    & 4,576  & 227   & 4,349 \\
          &       &       &       & Microaneurysms & 54    & 9,504  & 1,622  & 7,882 \\
          &       &       &       & Haemorrhages & 53    & 9,328  & 1,452  & 7,876 \\
          &       &       & N/A  & Optic Disc & 54    & -     & -     & - \\
    \bottomrule
    \end{tabular}}
    \vspace{-1em}
  \caption{\small Patch generation and selection for all datasets and features.}
  \label{tab:traininsummary}%
\end{table}%

\subsection{Retinal Datasets}
To assess the effectiveness of LMBF-Net, we performed segmentation experiments on five distinct retinal image datasets (Table \ref{tab:Datasets}). The DRIVE, STARE, CHASE and HRF datasets were used for retinal blood vessel segmentation, while IDRiD was used for microaneurysm, haemorrhages, soft and hard exudates and optic disc segmentation.

\subsection{Patch Generation, Selection, and Augmentation}
Insufficient pixel-label annotations are a significant obstacle in the analysis of retinal images. Limited data are the primary cause of overfitting in deep learning models. To prevent our model from overfitting, we used a patch-based implementation of LMBF-Net to segment DR lesions and blood vessels. Patches were generated from input images and selected or deleted based on whether they contained pathology pixels (Table \ref{tab:traininsummary}), and finally augmented using various transformations.

\subsection{Performance Measures}
The segmentation results were quantitatively evaluated by comparing them with the respective ground-truth images. The pixels were classified into correctly segmented foreground pixels (true positives $T_{P}$) and background pixels (true negatives $T_{N}$) and incorrectly segmented foreground pixels (false positives $F_{P}$) and background pixels (false negatives $F_{N}$). From these, the accuracy ($A_{cc}$), sensitivity ($S_{n}$), and the specificity ($S_{p}$), $F_{1}$ score, and \emph{AUC} were computed (defined in \cite{iqbal2022recent}).

\subsection{Ablation Study}
\begin{figure}[!t]
\centering
    \includegraphics[width=\columnwidth]{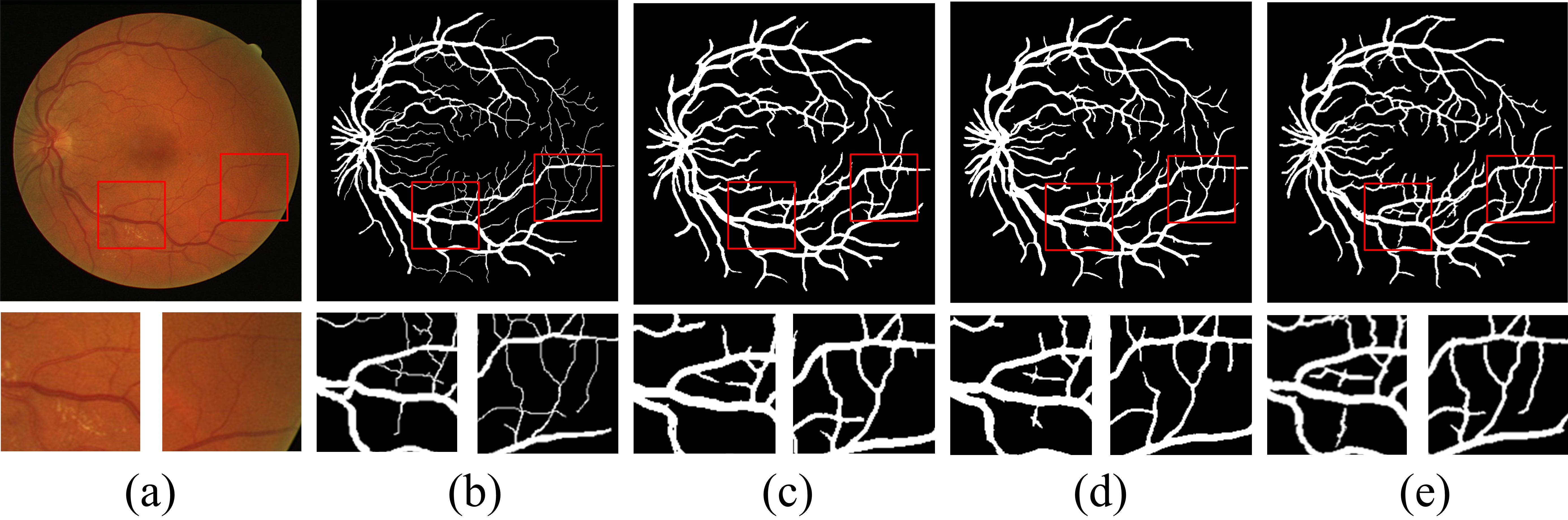}
    \vspace{-2.5em}
    \caption{\small Illustration of the ablation analysis conducted on the DRIVE dataset for retinal vascular segmentation. (a) RGB image input. (b) Corresponding ground truth. (c) Baseline network Bio-Net output. (d) Bio-Net++ output. (e) LMBF-Net output.}
    \label{fig: Ablation}
\end{figure}

The segmentation output of the LMBF-Net exhibits a higher degree of consistency with the ground truth when compared to Bio-Net and Bio-Net++ as can be easily observed from the results of an ablation study (Table \ref{tab:Ablation} and Fig.~\ref{fig: Ablation}). Tiny vessels that were completely missed by Bio-Net and partially missed by Bio-Net++ were almost completely captured by the proposed method, thus improving its performance. The enhancement in performance can be attributed to the utilization of multiscale feature information propagation and focal modulation between the encoder and decoder.

The training of the proposed LMBF-Net converged almost 2.7 times faster than that of Bio-Net. This is because employing the optimal number of filters prevents filter overlap and helps the network converge significantly faster than Bio-Net. For example, LMBF-Net took only 36 epochs to reach the point that Bio-Net \cite{xiang2020bio} reached after 100 epochs.

\begin{figure}[!t]
  \centering
  \resizebox{1\textwidth}{!}{%
  \begin{tabular}{@{}ccccc@{}}
       \includegraphics[width=1\textwidth]{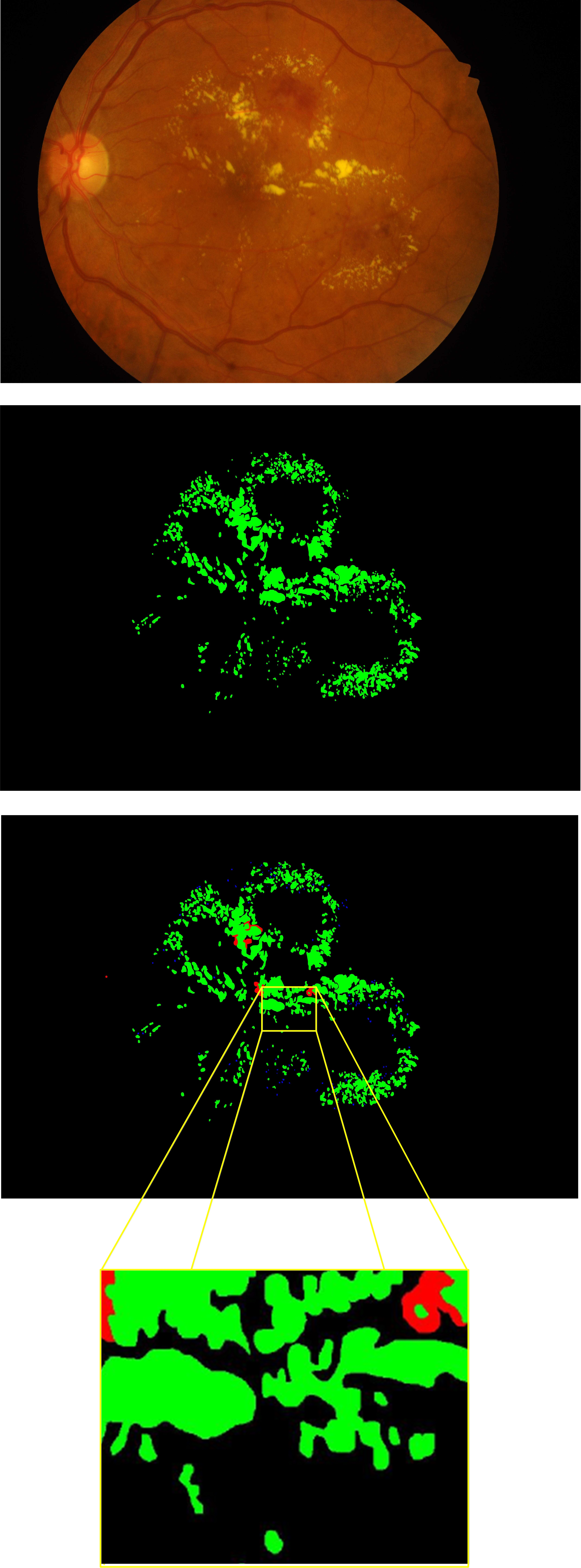} &  \includegraphics[width=1\textwidth]{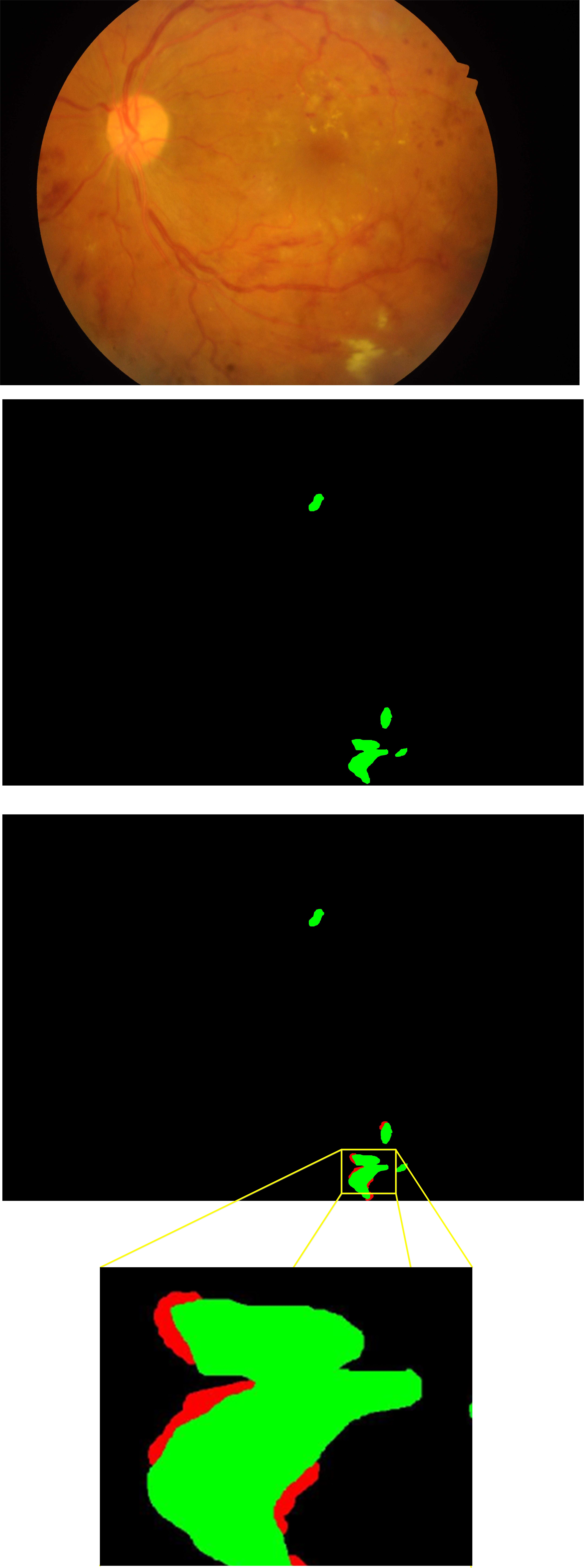} &  \includegraphics[width=1\textwidth]{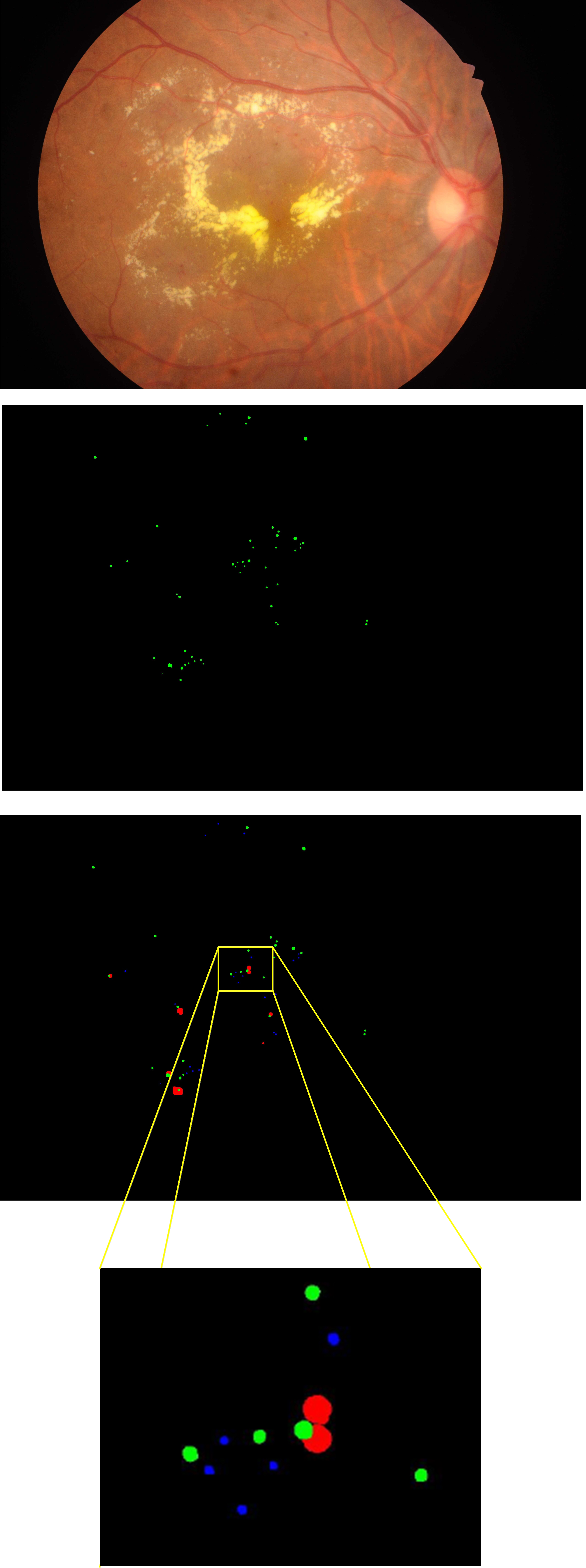} &  \includegraphics[width=1\textwidth]{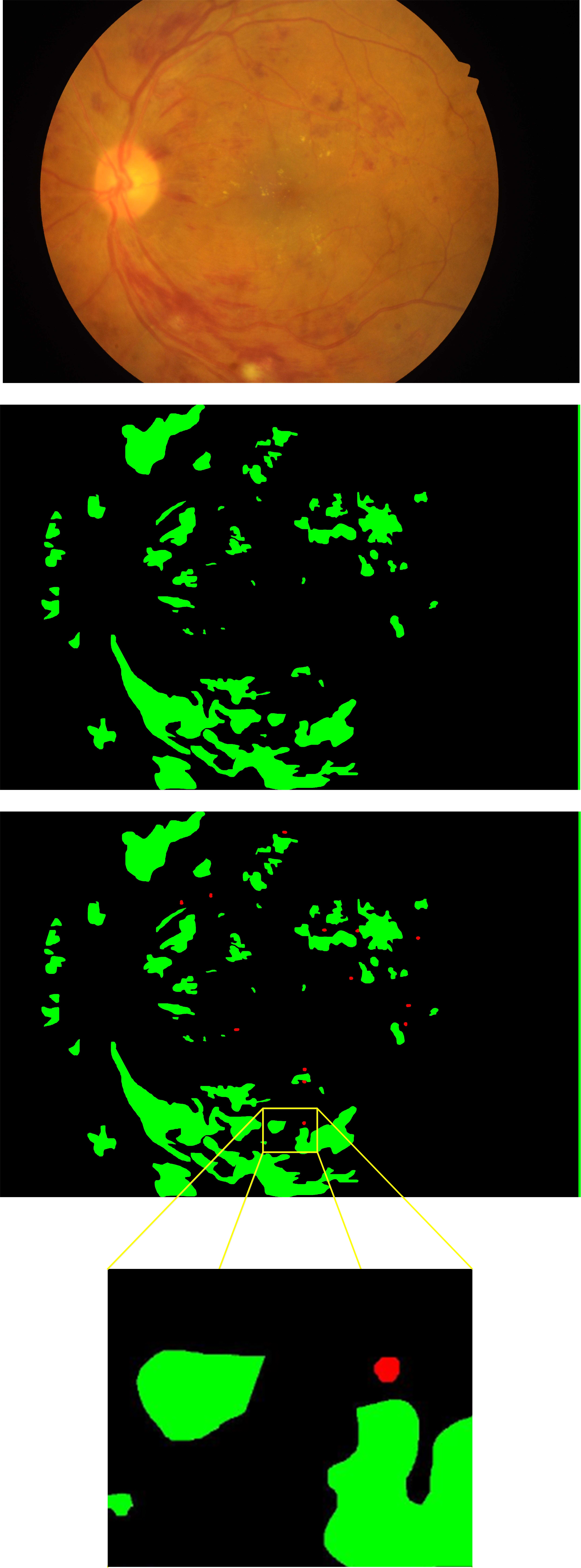} &  \includegraphics[width=1\textwidth]{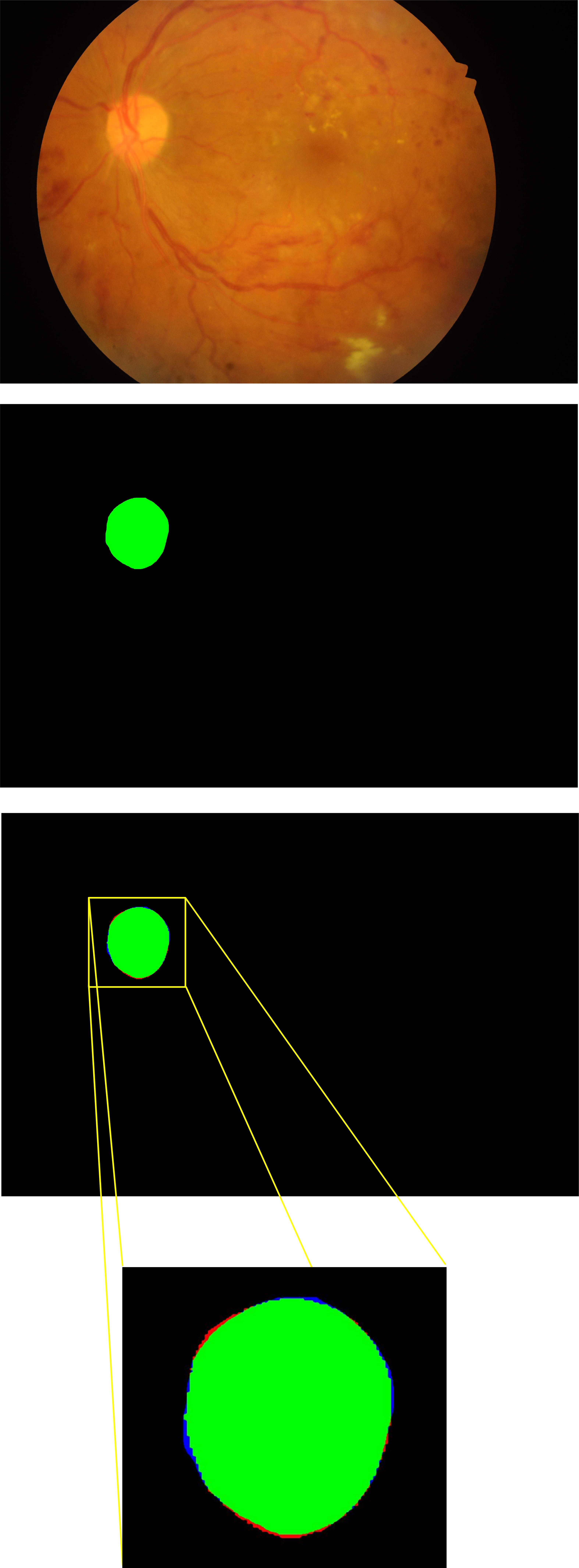} \\
  \end{tabular}}
    \vspace{-1em}
    \caption{\small Segmentation results of our LMBF-Net on representative test images from the IDRiD dataset. Top to bottom row: input images, ground truth, LMBF-Net results, and zoomed views of selected regions. Left to right column: hard exudates, soft exudates, microaneurysms, haemorrhages, and optic disc. False positive pixels are coloured red, while blue pixels show false negatives.}
  \label{fig:LesionVisual}
\end{figure}

\begin{figure}[!t]
\centering
  \includegraphics[width=\columnwidth]{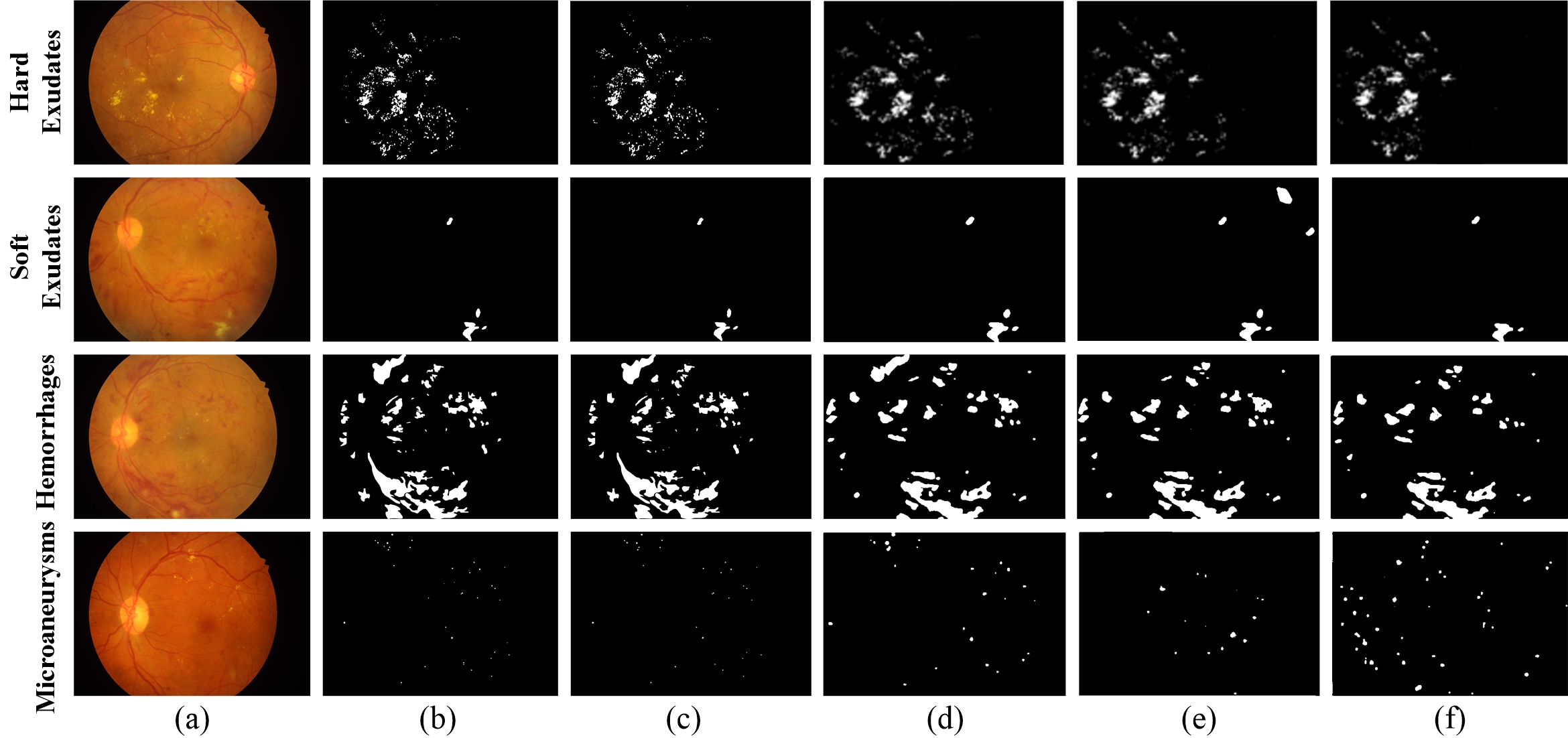}
  \caption{\small Visual comparison of sample segmentation results of the proposed LMBF-Net and recent networks on the IDRiD dataset. (a) Input RGB images. (b) Corresponding ground truths. (c) Results of LMBF-Net. (d) Results of CARNet \cite{guo2022carnet}. (e) Results of Bio-Net \cite{xiang2020bio}. (f) Results of \cite{foo2020multi}.}
    \vspace{-1em}
  \label{fig:LesionComparison}
\end{figure}

\begin{table*}[htbp]
  \centering
   \adjustbox{max width=\textwidth}{%
    \begin{tabular}{lcccccc}
    \toprule
    \multirow{2}[4]{*}{\textbf{Method}} & \multicolumn{1}{r}{\multirow{2}[4]{*}{\textbf{\#Params (M)}}} & \multicolumn{5}{c}{\textbf{Performance (\%)}} \\
\cmidrule{3-7}          &       & $S_{n}$ & $S_{p}$ & $A_{cc}$ & $F_{1}$ & \emph{AUC} \\
    \midrule
    Baseline (Bio-Net) \cite{xiang2020bio} & 14.99 & 0.8220 & 0.9804 & 0.9609 & 0.8206 & 0.9826 \\
    Bio-Net++ \cite{xiang2020bio} & 1.4 & 0.8363 & 0.9847 & 0.9652 & 0.8349 & 0.9839 \\
    Baseline with reduced parameters (BRP) & 0.092 & 0.7842 & 0.9741 & 0.9580 & 0.8086 & 0.9388 \\
    BRP + MRB [$f^{(1,1)}$+$f^{(3,3)}$] & 0.095 & 0.8207 & 0.9825 & 0.9682 & 0.8199 & 0.9802 \\
    BRP + MRB [$f^{(1,1)}$+$f^{(3,3)}$+$f^{(5,5)}$] & \multirow{3}[0]{*}{0.155} & 0.8249 & 0.9828 & 0.9689 & 0.8224 & 0.9845 \\
    BRP + MRB [$f^{(1,1)}$+$f^{(3,3)}$+$f^{(5,5)}$] + Patch-wise ($64 \times 64$) &       & 0.8148 & 0.9742 & 0.9596 & 0.8225 & 0.9820 \\
    BRP + MRB [$f^{(1,1)}$+$f^{(3,3)}$+$f^{(5,5)}$] + Patch-wise ($128 \times 128$) &       & 0.8318 & 0.9777 & 0.9617 & 0.8381 & 0.9810 \\
    \textbf{BRP + MRB [$f^{(1,1)}$+$f^{(3,3)}$+$f^{(5,5)}$] + Patch-wise ($128 \times 128$) + FMAB} & \textbf{0.191} & \textbf{0.8443} & \textbf{0.9861} & \textbf{0.9697} & \textbf{0.8459} & \textbf{0.9846} \\
    \bottomrule
    \end{tabular}}
  \caption{\small Results of the ablation study for retinal vessels segmentation performed on the DRIVE dataset.} 
    \vspace{-1em}
  \label{tab:Ablation}
\end{table*}

\subsection{Comparison With Recent Methods}
We also performed quantitative and qualitative comparisons of LMBF-Net with other recent methods. In the literature, the quantitative performance on DRIVE, STARE, CHASE, HRF, and IDRiD is typically reported in terms of $S_{n}$, $S_{p}$, $A_{cc}$, $F_{1}$, and \emph{AUC}. For comparison with other networks, we copied their performance scores from the cited papers if available (unknown results are indicated by a dash).

\begin{table*}[!t]
  \centering
   \adjustbox{max width=\textwidth}{%
    \begin{tabular}{lcccccccccccccccccc}
    \toprule
    \multirow{3}[6]{*}{\textbf{Method}} & \multirow{3}[6]{*}{\textbf{\#Params (M)}} & \multicolumn{17}{c}{\textbf{Performance (\%)}} \\
\cmidrule{3-19} & & \multicolumn{5}{c}{\textbf{DRIVE}} & & \multicolumn{5}{c}{\textbf{STARE}} &  & \multicolumn{5}{c}{\textbf{CHASE}} \\
\cmidrule{3-7}\cmidrule{9-13}\cmidrule{15-19} & & $S_{n}$ & $S_{p}$ & $A_{cc}$ & $AUC$ & $F_{1}$ & & $S_{n}$ & $S_{p}$ & $A_{cc}$ & \emph{AUC} & $F_{1}$ & & $S_{n}$ & $S_{p}$ & $A_{cc}$ & \emph{AUC} & $F_{1}$ \\
    \midrule
    UNet++ \cite{zhou2018unet++}  & 9.04 & - & \textbf{98.93} & 97.92 & 96.11 & 81.35 & & 79.01 &  76.80 & 97.38 & 95.19 & 82.70 &       & 81.12 & 98.24 & 97.02 & 98.42 & 81.22 \\
    Att UNet \cite{oktay2018attention} & 34.86 & 79.46 & 97.89 & 95.64 & 97.99 & 82.32 &       & 77.09 & 98.48 & 96.33 & 97.00 & -     &       & 80.10 & 98.04 & 96.42 & 98.40 & 80.12 \\
    BTS-DSN \cite{Guo2019} & 7.80 & 78.00 & 98.06 & 95.51 & 97.96 & 82.08 &       & 82.01 & 98.28 & 96.60 & 98.72 & 83.62 &       & 78.88 & 98.01 & 96.27 & 98.40 & 79.83 \\
    DUNet \cite{Jin2019} & 0.90 & 79.84 & 98.03 & 95.75 & 98.11 & 82.49 &       & 78.92 & 98.16 & 96.34 & 98.43 & 82.30 &       & 77.35 & 98.01 & 96.18 & 98.39 & 79.32 \\
    M2U-Net \cite{laibacher2019m2u} & 0.55 & 79.80    & 97.99    & 96.30 & 97.14 & 80.91 &       & 78.90 & 98.01 & 96.22 & 98.33 & 82.12 &       & 80.15 & 98.11 & 97.03 & 96.66 & 80.06 \\
    Baseline (Bio-Net) \cite{xiang2020bio} & 14.99 & 82.20 & 98.04 &96.09 & 98.26 & 82.06 &       & 81.70 & 98.06 & 96.82 & 98.83 & 82.22 &       & 82.21 & 98.24 & 96.32 & 98.10 & 80.02 \\
    CC-Net \cite{Feng2020} & -     & 76.25 & 98.09 & 95.28 & 96.78 & -     &       & 80.67 & 98.16 & 96.32 & 98.33 & 81.36 &       & -     & -     & -     & -     & - \\
    Wave-Net \cite{liu2022wave} & 1.50 & 81.64 & 97.64 & 95.61 & -     & 82.54 &       & 79.02 & 98.36 & 96.41 & -     & 81.40 &       & 82.83 & 98.21 & 96.64 & -     & \textbf{83.49} \\
    \textbf{Proposed LMBF-Net} & \textbf{0.19} & \textbf{83.48} & 98.77 & \textbf{96.97} & \textbf{98.46} & \textbf{83.81} &       & \textbf{85.19} & \textbf{98.61} & \textbf{97.59} & \textbf{99.06} & \textbf{84.22} &       & \textbf{83.29} & \textbf{98.34} & \textbf{97.31} & \textbf{98.82} & 82.10 \\
    \bottomrule
    \end{tabular}}
    \vspace{-1em}
  \caption{\small Performance comparison of LMBF-Net with recent methods on the DRIVE, STARE, and CHASE datasets. Best results are in bold. The performance of other methods is quoted from the cited papers. Unreported results are indicated by a dash.}
  \label{tab:DRIVE}
\end{table*}

\begin{table*}[!t]
  \centering
   \adjustbox{max width=\textwidth}{%
    \begin{tabular}{lccccccccccccccccccc}
    \toprule
    \multirow{3}[6]{*}{\textbf{Method}} & \multicolumn{19}{c}{\textbf{Performance (\%)}} \\
\cmidrule{2-20}          & \multicolumn{4}{c}{\textbf{Hard Exudates}} &       & \multicolumn{4}{c}{\textbf{Soft Exudates}} &       & \multicolumn{4}{c}{\textbf{Microaneurysms}} &       & \multicolumn{4}{c}{\textbf{Hemorrhages}} \\
\cmidrule{2-5}\cmidrule{7-10}\cmidrule{12-15}\cmidrule{17-20}          & \textbf{$S_{n}$} & \textbf{$S_{p}$} & \textbf{$A_{cc}$} & \textbf{$F_{1}$} &       & \textbf{$S_{n}$} & \textbf{$S_{p}$} & \textbf{$A_{cc}$} & \textbf{$F_{1}$} &       & \textbf{$S_{n}$} & \textbf{$S_{p}$} & \textbf{$A_{cc}$} & \textbf{$F_{1}$} &       & \textbf{$S_{n}$} & \textbf{$S_{p}$} & \textbf{$A_{cc}$} & \textbf{$F_{1}$} \\
    \midrule
    Bio-Net \cite{xiang2020bio} & 87.58 & 98.73 & 86.32 & 85.39 &       & 81.47 & 82.63 & 87.59 & 92.46 &       & 46.06 & 95.63 & 85.42 & 76.97 &       & 62.57 & \textbf{98.93} & 94.52 & 91.37 \\
    Deep Residual Network \cite{mo2018exudate} & 96.30 & 93.04 & 94.08 & 95.17 &       & 86.97 & 93.71 & 93.84 & 92.67 &       & 81.52 & 79.86 & 91.26 & 90.42 &       & 85.63 & 83.47 & 76.98 & 91.56 \\
    L-SegNet \cite{guo2019seg} & 81.65 & 76.13 & 85.64 & 89.71 &       & 78.69 & 81.54 & 83.52 & 87.93 &       & 61.72 & 54.63 & 52.49 & 46.57 &       & 63.74 & 65.81 & 61.28 & 59.87 \\
    CNN-based model \cite{foo2020multi} & 85.73 & 85.62 & 90.17 & 88.93 &       & 86.97 & 84.53 & 91.24 & 84.27 &       & 73.14 & 72.59 & 79.86 & 75.15 &       & 81.69 & 82.73 & 82.14 & 75.92 \\
    Modified U-Net \cite{sambyal2020modified} & 93.35 & 94.82 & 95.74 & 96.81 &       & 94.52 & 96.13 & 94.87 & 95.46 &       & 84.58 & 90.17 & 89.45 & 87.28 &       & 81.26 & 79.85 & 82.63 & 81.71 \\
    CARNet \cite{guo2022carnet} & 96.27 & 98.52 & 98.23 & \textbf{97.82} &       & 94.57 & 96.85 & 97.52 & 96.42 &       & 89.36 & 93.28 & 92.57 & 91.83 &       & 93.82 & 95.21 & 95.37 & \textbf{94.78} \\
    \textbf{Proposed LMBF-Net} & \textbf{96.34} & \textbf{99.70} & \textbf{99.63} & 95.00 &       & \textbf{99.52} & \textbf{99.94} & \textbf{99.92} & \textbf{99.28} &       & \textbf{94.57} & \textbf{99.90} & \textbf{99.87} & \textbf{93.28} &       & \textbf{95.12} & \textbf{98.33} & \textbf{98.26} & 92.62 \\
    \bottomrule
    \end{tabular}}
  \caption{\small Performance comparison of LMBF-Net with other methods on the IDRiD dataset for the segmentation of hard exudates, soft exudates, microaneurysms, and haemorrhages. Best results are in bold. The performance of other methods is quoted from the cited papers.}
    \vspace{-1em}
  \label{tab:IDRiD}
\end{table*}

\begin{table}[!t]
  \centering\footnotesize
    \begin{tabular}{lcc}
    \hline
    \textbf{Method} & \textbf{Parameters (M)} & \textbf{Inference Time (ms)} \\
    \hline
    M2U-Net \cite{laibacher2019m2u} & 0.550 & 20.25 \\
    Att UNet \cite{oktay2018attention} & 34.860 & 36.89 \\
    UNet++ \cite{zhou2018unet++} & 9.040 & 25.22 \\
    Baseline (Bio-Net) \cite{xiang2020bio} & 14.990 & 26.53 \\ 
    CE-Net \cite{gu2019net} & 29.00    & 24.03 \\
    Wave-Net \cite{liu2022wave} & 1.50   & 29.36 \\
    \textbf{LMBF-Net}  & \textbf{0.191} & \textbf{17.86} \\
    \hline
    \end{tabular}
    \vspace{-1em}
  \caption{\small Computational complexity of LMBF-Net and alternative segmentation networks.}
  \label{tab:performanceComparisons}
\end{table}

From the quantitative comparison with other recent methods for blood vessel segmentation on DRIVE, STARE, and CHASE (Table~\ref{tab:DRIVE}), it is evident that the proposed LMBF-Net outperformed all other methods and consistently achieved the highest score for all metrics. At the same time, LMBF-Net is one of the smallest networks in terms of the number of learnable parameters. That is, the proposed network achieved the best performance on all datasets with only 0.191M parameters. 

Another key factor in the evaluation of networks is their computational complexity in terms of trainable parameters and inference time. Comparing recent vessel segmentation architectures such as M2U-Net \cite{laibacher2019m2u}, Att UNet \cite{oktay2018attention}, UNet++ \cite{zhou2018unet++}, Bio-Net \cite{xiang2020bio}, CE-Net \cite{gu2019net}, and Wave-Net \cite{liu2022wave} with LMBF-Net on the DRIVE dataset (Table \ref{tab:performanceComparisons}), we see that LMBF-Net is smaller than the other architectures (only 0.191M learnable parameters) and also has the minimum inference time (17.86 ms). This makes the proposed network a better choice for real-time clinical implementation.

These quantitative results are confirmed by visual inspection of the LMBF-Net sample segmentation results (Figs.~\ref{fig:LesionVisual},\ref{fig:LesionComparison}).

\section{Conclusions}
\label{sec:Conclusions}

This study presents a medical image segmentation model named the Lightweight Multipath Bidirectional Focal Attention Network (LMBF-Net), which comprises a mere 0.191M learnable parameters. LMBF-Net is designed with a combination of convolution and group convolution operations, which significantly improves the computational efficiency of the network. The proposed network uses an optimal number of filters to prevent feature overlap and converges much faster than Bio-Net. This significantly reduces training time. The proposed network incorporates focal modulation between the encoder and the decoder for the refinement of encoder features. A patch selection strategy is also proposed to solve the class imbalance issue of medical image segmentation. The experimental findings of LMBF-Net on various features of the retinal image, such as retinal vessels, hard exudates, soft exudates, haemorrhages, microaneurysms, and optic disc, confirm the robustness and versatility of the proposed network.
LMBF-Net outperforms other segmentation methods for retinal vessels and DR lesions.


\begin{thebibliography}{10}

\bibitem{iqbal2022recent}
S.~Iqbal, T.~M. Khan, K.~Naveed, S.~S. Naqvi, and S.~J. Nawaz,
\newblock ``{Recent trends and advances in fundus image analysis: A review},''
\newblock {\em Computers in Biology and Medicine}, vol. 151, pp. 106277, 2022.

\bibitem{franklin2014computerized}
S.~W. Franklin and S.~E. Rajan,
\newblock ``{Computerized screening of diabetic retinopathy employing blood
  vessel segmentation in retinal images},''
\newblock {\em Biocybernetics and Biomedical Engineering}, vol. 34, no. 2, pp.
  117--124, 2014.

\bibitem{iqbal2023robust}
S.~Iqbal, K.~Naveed, S.~S. Naqvi, A.~Naveed, and T.~M. Khan,
\newblock ``Robust retinal blood vessel segmentation using a patch-based
  statistical adaptive multi-scale line detector,''
\newblock {\em Digital Signal Processing}, vol. 139, pp. 104075, 2023.

\bibitem{manan2023semantic}
M.~A. Manan, F.~Jinchao, T.~M. Khan, M.~Yaqub, S.~Ahmed, and I.~S. Chuhan,
\newblock ``{Semantic segmentation of retinal exudates using a residual
  encoder--decoder architecture in diabetic retinopathy},''
\newblock {\em Microscopy Research and Technique}, vol. 86, no. 11, pp.
  1443--1460, 2023.

\bibitem{abbasi2023lmbis}
M.~M. Abbasi, S.~Iqbal, A.~Naveed, T.~M. Khan, S.~S. Naqvi, and W.~Khalid,
\newblock ``{LMBiS-Net: A Lightweight Multipath Bidirectional Skip Connection
  based CNN for Retinal Blood Vessel Segmentation},''
\newblock {\em arXiv preprint arXiv:2309.04968}, 2023.

\bibitem{khalid2023advancing}
W.~Khalid, M.~Y. Khalid, M.~Hena, A.~Sarwar, and S.~Iqbal,
\newblock ``{Advancing Pharmaceuticals with Machine Learning: A Short Review of
  Research and Development Applications},''
\newblock {\em Pharmaceutical Communications}, vol. 2, no. 01, pp. 63--69,
  2023.

\bibitem{khan2023retinal}
T.~M. Khan, S.~S. Naqvi, A.~Robles-Kelly, and I.~Razzak,
\newblock ``{Retinal vessel segmentation via a Multi-resolution Contextual
  Network and adversarial learning},''
\newblock {\em Neural Networks}, vol. 165, pp. 310--320, 2023.

\bibitem{long2015fully}
J.~Long, E.~Shelhamer, and T.~Darrell,
\newblock ``Fully convolutional networks for semantic segmentation,''
\newblock in {\em IEEE Conference on Computer Vision and Pattern Recognition
  (CVPR)}, 2015, pp. 3431--3440.

\bibitem{chen2021transunet}
J.~Chen, Y.~Lu, Q.~Yu, X.~Luo, E.~Adeli, Y.~Wang, L.~Lu, A.~L. Yuille, and
  Y.~Zhou,
\newblock ``{TransUNet}: Transformers make strong encoders for medical image
  segmentation,''
\newblock {\em arXiv:2102.04306}, 2021.

\bibitem{khan2024esdmr}
T.~M. Khan, S.~S. Naqvi, and E.~Meijering,
\newblock ``{ESDMR-Net: A lightweight network with expand-squeeze and dual
  multiscale residual connections for medical image segmentation},''
\newblock {\em Engineering Applications of Artificial Intelligence}, vol. 133,
  pp. 107995, 2024.

\bibitem{gu2019net}
Z.~Gu, J.~Cheng, H.~Fu, K.~Zhou, H.~Hao, Y.~Zhao, T.~Zhang, S.~Gao, and J.~Liu,
\newblock ``{CE-Net}: Context encoder network for {2D} medical image
  segmentation,''
\newblock {\em IEEE Transactions on Medical Imaging}, vol. 38, no. 10, pp.
  2281--2292, 2019.

\bibitem{iqbal2023ldmres}
S.~Iqbal, T.~M. Khan, S.~S. Naqvi, A.~Naveed, M.~Usman, H.~A. Khan, and
  I.~Razzak,
\newblock ``{LDMRes-Net: A Lightweight Neural Network for Efficient Medical
  Image Segmentation on IoT and Edge Devices},''
\newblock {\em IEEE Journal of Biomedical and Health Informatics}, 2023.

\bibitem{iqbal2022g}
S.~Iqbal, S.~S. Naqvi, H.~A. Khan, A.~Saadat, and T.~M. Khan,
\newblock ``{G-Net Light}: A lightweight modified {Google Net} for retinal
  vessel segmentation,''
\newblock {\em Photonics}, vol. 9, no. 12, pp. 923, 2022.

\bibitem{naqvi2023glan}
S.~S. Naqvi, Z.~A. Langah, H.~A. Khan, M.~I. Khan, T.~Bashir, M.~I. Razzak, and
  T.~M. Khan,
\newblock ``{GLAN: Gan assisted lightweight attention network for biomedical
  imaging based diagnostics},''
\newblock {\em Cognitive Computation}, vol. 15, no. 3, pp. 932--942, 2023.

\bibitem{khan2022mkis}
T.~M. Khan, M.~Arsalan, A.~Robles-Kelly, and E.~Meijering,
\newblock ``{MKIS-Net}: A light-weight multi-kernel network for medical image
  segmentation,''
\newblock {\em arXiv:2210.08168}, 2022.

\bibitem{li2021pyconvu}
C.~Li, Y.~Fan, and X.~Cai,
\newblock ``{PyConvU-Net}: A lightweight and multiscale network for biomedical
  image segmentation,''
\newblock {\em BMC Bioinformatics}, vol. 22, no. 1, pp. 1--11, 2021.

\bibitem{iqbal2023mlr}
S.~Iqbal, T.~M. Khan, S.~S. Naqvi, and G.~Holmes,
\newblock ``{MLR-Net: A multi-layer residual convolutional neural network for
  leather defect segmentation},''
\newblock {\em Engineering Applications of Artificial Intelligence}, vol. 126,
  pp. 107007, 2023.

\bibitem{liu2021swin}
Z.~Liu, Y.~Lin, Y.~Cao, H.~Hu, Y.~Wei, Z.~Zhang, S.~Lin, and B.~Guo,
\newblock ``Swin transformer: Hierarchical vision transformer using shifted
  windows,''
\newblock in {\em IEEE/CVF International Conference on Computer Vision (ICCV)},
  2021, pp. 10012--10022.

\bibitem{yang2022focal}
J.~Yang, C.~Li, X.~Dai, and J.~Gao,
\newblock ``Focal modulation networks,''
\newblock {\em Advances in Neural Information Processing Systems}, vol. 35, pp.
  4203--4217, 2022.

\bibitem{khan2020semantically}
T.~M. Khan, A.~Robles-Kelly, and S.~S. Naqvi,
\newblock ``A semantically flexible feature fusion network for retinal vessel
  segmentation,''
\newblock in {\em International Conference on Neural Information Processing
  (ICONIP)}, 2020, pp. 159--167.

\bibitem{mehmood2024retinalitenet}
M.~Mehmood, M.~Alsharari, S.~Iqbal, I.~Spence, and M.~Fahim,
\newblock ``{RetinaLiteNet: A Lightweight Transformer based CNN for Retinal
  Feature Segmentation},''
\newblock in {\em Proceedings of the IEEE/CVF Conference on Computer Vision and
  Pattern Recognition}, 2024, pp. 2454--2463.

\bibitem{khan2023feature}
T.~M. Khan, M.~Arsalan, S.~Iqbal, I.~Razzak, and E.~Meijering,
\newblock ``{Feature Enhancer Segmentation Network (FES-Net) for Vessel
  Segmentation},''
\newblock in {\em 2023 International Conference on Digital Image Computing:
  Techniques and Applications (DICTA)}. IEEE, 2023, pp. 160--167.

\bibitem{khan2022t}
T.~M. Khan, A.~Robles-Kelly, and S.~S. Naqvi,
\newblock ``{T-Net}: A resource-constrained tiny convolutional neural network
  for medical image segmentation,''
\newblock in {\em IEEE/CVF Winter Conference on Application of Computer Vision
  (WACV)}, 2022, pp. 644--653.

\bibitem{xiang2020bio}
T.~Xiang, C.~Zhang, D.~Liu, Y.~Song, H.~Huang, and W.~Cai,
\newblock ``{BiO-Net}: Learning recurrent bi-directional connections for
  encoder-decoder architecture,''
\newblock in {\em International Conference on Medical Image Computing and
  Computer-Assisted Intervention (MICCAI)}, 2020, pp. 74--84.

\bibitem{DRIVEdata}
J.~Staal, M.~D. Abr{\`a}moff, M.~Niemeijer, M.~A. Viergever, and
  B.~Van~Ginneken,
\newblock ``{Ridge-based vessel segmentation in color images of the retina},''
\newblock {\em IEEE Transactions Medical Imaging}, vol. 23, no. 4, pp.
  501--509, 2004.

\bibitem{STAREDataset}
A.~Hoover, V.~Kouznetsova, and M.~Goldbaum,
\newblock ``{Locating blood vessels in retinal images by piecewise threshold
  probing of a matched filter response},''
\newblock {\em IEEE Transactions Medical Imaging}, vol. 19, no. 3, pp.
  203--210, 2000.

\bibitem{CHASEDataset}
M.~M. Fraz, P.~Remagnino, A.~Hoppe, B.~Uyyanonvara, A.~R. Rudnicka, C.~G. Owen,
  and S.~A. Barman,
\newblock ``An ensemble classification-based approach applied to retinal blood
  vessel segmentation,''
\newblock {\em IEEE Transactions on Biomedical Engineering}, vol. 59, no. 9,
  pp. 2538--2548, 2012.

\bibitem{HRFDataset}
J.~Odstrcilik, R.~Kolar, A.~Budai, J.~Hornegger, J.~Jan, J.~Gazarek, T.~Kubena,
  P.~Cernosek, O.~Svoboda, and E.~Angelopoulou,
\newblock ``{Retinal vessel segmentation by improved matched filtering:
  evaluation on a new high-resolution fundus image database},''
\newblock {\em IET Image Processing}, vol. 7, no. 4, pp. 373--383, 2013.

\bibitem{IDRiDDataset}
P.~Porwal, S.~Pachade, R.~Kamble, M.~Kokare, G.~Deshmukh, V.~Sahasrabuddhe, and
  F.~Meriaudeau,
\newblock ``{Indian Diabetic Retinopathy Image Dataset {(IDRiD)}}: A database
  for diabetic retinopathy screening research,''
\newblock {\em Data}, vol. 3, no. 3, pp. 25, 2018.

\bibitem{guo2022carnet}
Y.~Guo and Y.~Peng,
\newblock ``{CARNet: Cascade attentive RefineNet for multi-lesion segmentation
  of diabetic retinopathy images},''
\newblock {\em Complex \& Intelligent Systems}, vol. 8, no. 2, pp. 1681--1701,
  2022.

\bibitem{foo2020multi}
A.~Foo, W.~Hsu, M.~L. Lee, G.~Lim, and T.~Y. Wong,
\newblock ``Multi-task learning for diabetic retinopathy grading and lesion
  segmentation,''
\newblock in {\em AAAI Conference on Artificial Intelligence}, 2020, pp.
  13267--13272.

\bibitem{zhou2018unet++}
Z.~Zhou, M.~M. Rahman~Siddiquee, N.~Tajbakhsh, and J.~Liang,
\newblock ``{UNet++}: A nested u-net architecture for medical image
  segmentation,''
\newblock in {\em International Workshops on Deep Learning in Medical Image
  Analysis (DLMIA) and Multimodal Learning for Clinical Decision Support
  (ML-CDS) Held in Conjunction with MICCAI}, 2018, pp. 3--11.

\bibitem{oktay2018attention}
O.~Oktay, J.~Schlemper, L.~L. Folgoc, M.~Lee, M.~Heinrich, K.~Misawa, K.~Mori,
  S.~McDonagh, N.~Y. Hammerla, B.~Kainz, B.~Glocker, and D.~Rueckert,
\newblock ``{Attention U-Net}: Learning where to look for the pancreas,''
\newblock {\em arXiv:1804.03999}, 2018.

\bibitem{Guo2019}
G.~Song, W.~Kai, K.~Hong, Z.~Yujun, G.~Yingqi, and L.~Tao,
\newblock ``{BTS-DSN}: Deeply supervised neural network with short connections
  for retinal vessel segmentation,''
\newblock {\em International Journal of Medical Informatics}, vol. 126, pp.
  105--113, 2019.

\bibitem{Jin2019}
Q.~Jin, Z.~Meng, T.~D. Pham, Q.~Chen, L.~Wei, and R.~Su,
\newblock ``{DUNet: A deformable network for retinal vessel segmentation},''
\newblock {\em Knowledge-Based Systems}, vol. 178, pp. 149--162, 2019.

\bibitem{laibacher2019m2u}
T.~Laibacher, T.~Weyde, and S.~Jalali,
\newblock ``{M2U-Net}: Effective and efficient retinal vessel segmentation for
  resource-constrained environments,''
\newblock in {\em IEEE/CVF Conference on Computer Vision and Pattern
  Recognition Workshops}, 2019, pp. 1--10.

\bibitem{Feng2020}
S.~Feng, Z.~Zhuo, D.~Pan, and Q.~Tian,
\newblock ``{CcNet}: A cross-connected convolutional network for segmenting
  retinal vessels using multi-scale features,''
\newblock {\em Neurocomputing}, vol. 392, pp. 268--276, 2020.

\bibitem{liu2022wave}
Y.~Liu, J.~Shen, L.~Yang, H.~Yu, and G.~Bian,
\newblock ``{Wave-Net: A lightweight deep network for retinal vessel
  segmentation from fundus images},''
\newblock {\em Computers in Biology and Medicine}, vol. 152, pp. 106341, 2022.

\bibitem{mo2018exudate}
J.~Mo, L.~Zhang, and Y.~Feng,
\newblock ``{Exudate-based diabetic macular edema recognition in retinal images
  using cascaded deep residual networks},''
\newblock {\em Neurocomputing}, vol. 290, pp. 161--171, 2018.

\bibitem{guo2019seg}
S.~Guo, T.~Li, H.~Kang, N.~Li, Y.~Zhang, and K.~Wang,
\newblock ``{L-Seg: An end-to-end unified framework for multi-lesion
  segmentation of fundus images},''
\newblock {\em Neurocomputing}, vol. 349, pp. 52--63, 2019.

\bibitem{sambyal2020modified}
N.~Sambyal, P.~Saini, R.~Syal, and V.~Gupta,
\newblock ``{Modified U-Net architecture for semantic segmentation of diabetic
  retinopathy images},''
\newblock {\em Biocybernetics and Biomedical Engineering}, vol. 40, no. 3, pp.
  1094--1109, 2020.

\end{thebibliography}
\end{document}